\documentclass[aip,rsi,reprint,graphicx]{revtex4-1} % for checking your page length
\usepackage{amssymb}
\newcommand{\ud}{\mathrm{d}}
\usepackage{amsmath}
\usepackage{verbatim}
\usepackage{lineno}
\begin{document}

%\preprint{}

\title{Rayleigh scattering of linear alkylbenzene in large liquid scintillator detectors} %Title of paper



\author{Xiang Zhou}
\email[]{xiangzhou@whu.edu.cn}
%\thanks{}
\affiliation{Hubei Nuclear Solid Physics Key Laboratory, Key Laboratory of Artificial Micro- and Nano-structures of Ministry of Education, and School of Physics and Technology, Wuhan University, Wuhan 430072, China}
\author{Qian Liu}
\affiliation{School of Physics, University of Chinese Academy of Sciences, Beijing 100049, China}
\author{Michael Wurm}
\affiliation{Institute of Physics \& EC PRISMA, Johannes Gutenberg University, 55128 Mainz, Germany}
\author{Qingmin Zhang}
\affiliation{Department of Nuclear Science and Technology, School of Energy and Power Engineering, Xi'an Jiaotong University, Xi'an 710049, China}
\author{Yayun Ding}
\affiliation{Institute of High Energy Physics, Chinese Academy of Sciences, Beijing 100049, China}
\author{Zhenyu Zhang}
\affiliation{Hubei Nuclear Solid Physics Key Laboratory, Key Laboratory of Artificial Micro- and Nano-structures of Ministry of Education, and School of Physics and Technology, Wuhan University, Wuhan 430072, China}
\author{Yangheng Zheng}
\affiliation{School of Physics, University of Chinese Academy of Sciences, Beijing 100049, China}
\author{Li Zhou}
\affiliation{Institute of High Energy Physics, Chinese Academy of Sciences, Beijing 100049, China}
\author{Jun Cao}
\affiliation{Institute of High Energy Physics, Chinese Academy of Sciences, Beijing 100049, China}
\author{Yifang Wang}
\affiliation{Institute of High Energy Physics, Chinese Academy of Sciences, Beijing 100049, China}


\date{\today}

\begin{abstract}
Rayleigh scattering poses an intrinsic limit for the transparency of organic liquid scintillators. This work focuses on the Rayleigh scattering length of linear alkylbenzene (LAB), which will be used as the solvent of the liquid scintillator in the central detector of the Jiangmen Underground Neutrino Observatory. We investigate the anisotropy of the Rayleigh scattering in LAB, showing that the resulting Rayleigh scattering length will be significantly shorter than reported before. Given the same overall light attenuation, this will result in a more efficient transmission of photons through the scintillator, increasing the amount of light collected by the photosensors and thereby the energy resolution of the detector.   
\end{abstract}


\maketitle %\maketitle must follow title, authors, abstract and \pacs

% References should be done using the \cite and \label commands
\section{Introduction}
Due to the large value of the neutrino mixing angle $\theta_{13}$\cite{KAbe,Adamson,YAbe,AnPRL,AnNIM,Ahn,An2013,An2014,KAbe2014}, the neutrino mass hierarchy (i.e. the sign of $\Delta m^2_{31}$ or $\Delta m^2_{32}$) can be determined based on an interference pattern occurring in the oscillations of reactor antineutrinos\cite{Zhan2008,Zhan2009,Cahn}. The Jiangmen Underground Neutrino Observatory (JUNO) aims at a precision measurement of the reactor antineutrino spectrum at a medium baseline of 53 km from the source to resolve this subdominant effect\cite{Li2013,Li2014}. The central detector of JUNO is a large liquid-scintillator (LS) spherical detector with the diameter of about 35 m. The LS will be composed of linear alkylbenzene (LAB) acting as the solvent, 2,5-diphenyloxazole (PPO) as the fluor and 1,4-bis[2-methylstyry]benzene (bis-MSB) as the wavelength shifter. Electron antineutrinos are detected via the inverse beta decay reaction $\bar{\nu}_e + p \to e^+ + n$. To determine the neutrino mass hierarchy, the energy resolution of JUNO needs to be $3\%/\sqrt{E_{e^+}[\mathrm{MeV}]}$. This requires a minimum of 1200 photon electrons per MeV to be detected by the photomultipliers (PMTs)\cite{Li2013}. Since the scintillation photons generated in LS have to travel several tens of meters to the PMTs at the verge of the detector, high transparency of the LS solvent is of uttermost importance\cite{Cahn}. 

If multiple scattering is negligible, the numbers of photons in a beam of light will be exponentially attenuated from $N_0$ to $N$ after travelling a distance $x$ in a medium, that is 
\begin{displaymath}
N=N_0e^{-x/L},
\end{displaymath}
where $L$ is the attenuation length. The attenuation of the scintillation light is the result of the combination of absorption and scattering processes\cite{Bohren}, which can be described by the following formula,  
\begin{equation}
\frac{1}{L}=\frac{1}{L_\mathrm{abs}}+\frac{1}{L_\mathrm{sca}}, 
\label{abs}
\end{equation}
where $L_\mathrm{abs}$ and $L_\mathrm{sca}$ are the absorption length and scattering length, respectively. While absorption processes will convert photon excitations into heat and absorbed photons will thus be lost for detection, scattering will change only the travelling directions of photons. As they can still be collected by the PMTs in a 4$\pi$ detector, scattering does not degrade the energy resolution of the detector immediately. For a given attenuation length, a short scattering length is therefore preferable over a short absorption length in order to maximize the light collection. The effective wavelength is shifted to the blue range of the visible spectrum by the addition of PPO and bis-MSB to the LS. Therefore, self-absorption by LAB is not an issue\cite{Xiao}. The resulting spectral maximum of the scintillation light is is around 430 nm, which is usually chosen as the reference wavelength in the discussion of the light propagation in LS. Due to the importance of optical purity, the LAB used for JUNO will undergo a number of purification steps that will remove dust and organic impurities from the solvent that might otherwise feature absorption bands in the wavelength region of interest. Thus the light scattering in LAB will be dominated by Rayleigh scattering. 

A direct measurement of the absorption length of liquids is difficult. However, it can be indirectly determined by measuring the attenuation and scattering lengths and relying on Eq.~(\ref{abs})\cite{Querry}. For purified LS samples, lab experiments measuring the light attenuation over distances of several meters have found attenuation lengths on the order of 20 m at 430 nm\cite{Ding,Goett,Gao}. In addition, a series of laboratory-scale scattering experiments has been performed that included results for LAB at 430 nm\cite{Wurm}. It was found that only part of the scattered light amplitude follows the well-known angular distribution predicted by Rayleigh's theory\cite{Rayleigh1899,Alimonti}, $P(\theta)\propto(1+\cos^2\theta)$. The corresponding Rayleigh scattering length was determined to 40 m at 430 nm\cite{Wurm}. A remaining, isotropic component of light scattering in the LAB sample has been attributed to absorption/re-emission processes\cite{Wurm}. Assuming the above results on the attenuation length of 20 m, an absorption length of 40 m can be deduced from Eq.~(\ref{abs}), including the re-emission processes. However, it has been found both theoretically and experimentally that the angular distribution of Rayleigh scattering in liquids can deviate from $P(\theta)\propto(1+\cos^2\theta)$ because of the molecular anisotropy. Under this assumption, we re-examine the reported data of light scattering to obtain a corrected value for the scattering length and thereby the absorption length of LAB, which are crucial to determine the energy resolution of JUNO experiment. 

In this paper, we study Rayleigh scattering of LAB. Section \ref{sec2} gives the theory of Rayleigh scattering in liquids. Section \ref{sec3} gives the Rayleigh scattering length of LAB both from theory and the reported data. Section \ref{sec4} gives discussions and conclusions. 
%\label{}
\section{Theory of Rayleigh scattering in liquids\label{sec2}}
Rayleigh scattering can be characterized by the volume scattering function $\beta(\theta)$\cite{Jerlov},
\begin{equation}
\beta(\theta)=\frac{I(\theta)}{I_0}\frac{r^2}{V},
\end{equation}
where $I(\theta)$ is the intensity of scattered light in the direction $\theta$, $I_0$ the intensity of incident light, $r$ the distance from the scattering center and $V$ the scattering volume. The Rayleigh scattering length $l_\mathrm{Ray}$ is the reciprocal of the integral of the volume scattering function $\beta(\theta)$ over the solid angle 
\begin{equation}
l_\mathrm{Ray} = \left[\iint \beta(\theta)\ud\Omega\right]^{-1}.
\label{Rayintegral}
\end{equation}

The intensity of scattered light $I(\theta)$ can be divided into two parts in the scattering plane
\begin{equation}
I(\theta)=I_\parallel(\theta)+I_\perp(\theta),
\end{equation}
where $I_\parallel$ and $I_\perp$ are the parallel and perpendicular parts of the scattering intensity, respectively. The angular distributions of the scattering intensity generally are\cite{Dawson}
\begin{eqnarray}
I_\parallel(\theta) & = & \frac{\cos^2\theta}{2}A+\frac{1}{2}B,\nonumber\\
I_\perp(\theta) & = & \frac{1}{2}A+\frac{1}{2}B,
\label{aniso} 
\end{eqnarray}
where $A$ and $B$ are the measurable quantities. The polarization of scattered light at 90$^\circ$ can be characterized by the depolarization ratio $\delta$, which is
\begin{equation}
\delta=\frac{I_\parallel(90^\circ)}{I_\perp(90^\circ)}=\frac{B}{A+B}.
\label{delta}
\end{equation}

It is convenient to define the Rayleigh ratio $R$ as the volume scattering function $\beta(\theta)$ at 90$^\circ$,
\begin{equation}
R\equiv\beta(90^\circ)=\frac{I(90^\circ)}{I_0}\frac{r^2}{V}.
\end{equation}
Then the volume scattering function $\beta(\theta)$ of Rayleigh scattering is 
\begin{equation}
\beta(\theta) = R\left(1+\frac{1-\delta}{1+\delta}\cos^2\theta\right).
\label{beta}
\end{equation}
According to Eq. (\ref{Rayintegral}), the Rayleigh scattering length $l_\mathrm{Ray}$ can be determined by
\begin{equation}
l_\mathrm{Ray} = \left[\frac{8\pi}{3}R\frac{2+\delta}{1+\delta}\right]^{-1}.
\label{Raylength}
\end{equation}
Eq. (\ref{Raylength}) shows that Rayleigh scattering should be characterized by not only the Rayleigh scattering length or the Rayleigh ratio but also the depolarization ratio. 

For isotropic liquids the depolarization ratio is zero and the scattering length depends only on the Rayleigh ratio. Einstein\cite{Einstein} and Smoluchowski\cite{Smoluchowski} developed a phenomenological theory, which is applicable to the light scattering in condensed isotropic liquids. If the liquid is perfectly uniform, scattering will be completely eliminated by destructive interference except in the forward direction\cite{Jenkins}. However, liquids are usually not perfectly uniform. The local density of a liquid is constantly fluctuating because of the thermal motions of molecules. Since the local density fluctuations are random, the scattering off these fluctuations is incoherent and will increase the scattering amplitude. According to the Einstein-Smoluchowski theory, the Rayleigh ratio of isotropic liquids $R_\mathrm{iso}$ can be expressed by
\begin{equation}
R_\mathrm{iso}=\frac{\pi^2}{2\lambda^4}\left[\rho\left(\frac{\partial\varepsilon}{\partial\rho}\right)_T\right]^2kT\kappa_T,
\end{equation}
where $\lambda$ is the wavelength of the scattered light, $\rho$ the density of liquid, $\varepsilon$ the average dielectric constant of the liquid, $k$ the Boltzmann constant, $T$ the temperature, and $\kappa_T$ the isothermal compressibility.

Cabannes showed that the ratio of $R$ and $R_\mathrm{iso}$ satisfies\cite{Cabannes,Rayleigh1920a,Rayleigh1920b,King,Prins} 
\begin{equation}
\frac{R}{R_\mathrm{iso}}=\frac{6+6\delta}{6-7\delta},
\label{Cabannes}
\end{equation}
where $R_\mathrm{iso}$ can be treated as the isotropic part of the total Rayleigh ratio $R$. It is common to use empirical expressions to calculate $\rho(\partial\varepsilon/\partial\rho)_T$. For organic liquids, it was shown that the Eykman equation describes $\rho(\partial\epsilon/\partial\rho)$ well\cite{Carr,Shakparonov,Parfitt,Azim}. The Eykman equation is\cite{Eykman}
\begin{equation}
\rho\left(\frac{\partial\varepsilon}{\partial\rho}\right)_T=\frac{(n^2-1)(2n^2+0.8n)}{n^2+0.8n+1},
\end{equation}
where $n$ is the refractive index of liquid. Therefore, the Rayleigh scattering length of organic liquids can be determined by the so-called Einstein-Smoluchowski-Cabannes formula\cite{Coumou} 
\begin{equation}
l_\mathrm{Ray}=\left\{\frac{8\pi^3}{3\lambda^4}\left[\frac{(n^2-1)(2n^2+0.8n)}{n^2+0.8n+1}\right]^2kT\kappa_T\frac{6+3\delta}{6-7\delta}\right\}^{-1}.
\label{ESC}
\end{equation}

\section{Rayleigh scattering length and depolarization ratio for linear alkylbenzene\label{sec3}}
Based on the Einstein-Smoluchowski-Cabannes formula, the Rayleigh scattering length of LAB can be determined as a function of the incident wavelength $\lambda$ and the temperature $T$. The liquid parameters required are the isothermal compressibility $\kappa_T$, the refractive index $n$ and the depolarization ratio $\delta$. 

The isothermal compressibility $\kappa_T$ of LAB can been derived from its density equation of state $\rho=\rho(T,p)$ by 
\begin{equation}
\kappa_T=\frac{1}{\rho}\left(\frac{\partial\rho}{\partial p}\right)_T,
\end{equation}
where $p$ is the pressure. The density equation at constant temperature can be obtained by fitting the equation of state\cite{Holder}
\begin{equation}
\rho(p)=\rho_0(1+Ap+Bp^2).
\end{equation}
The density of LAB at 23$^\circ$C has been measured by a vibrating tube densimeter\cite{Yin} at 8 different pressures. The isothermal compressibility of LAB at 23$^{\circ}$C is 7.743$\pm$0.035$\times$10$^{-10}$Pa$^{-1}$ at normal pressure\cite{Zhou}.

The refractive index of LAB has been determined for 5 different wavelengths using the V-Prism refractometer at the National Institute of Metrology, China (Table~\ref{refractivity}). The dispersion can be parameterized by the Sellmeier equation 
\begin{equation}
n^2(\lambda) = 1+ \frac{B}{1-C/\lambda^2},
\end{equation}
where $B$ and $C$ are the fitting parameters. The refractive index of LAB is 1.49829$\pm$0.00026 at 430 nm at 19$^\circ$C. The change of the refractive index is less than 0.001 per degree Celsius\cite{Coumou}, so that 1.498$\pm$0.004 will hold as a good approximation for 23$^\circ$C. The results can be compared to measurements by the RENO collaboration using the minimum deviation technique. The refractive index was determined at six wavelengths in the range between 400 nm and 630 nm at room temperature\cite{Yeo}. By fitting the above formula to those results we obtain 1.49623$\pm$0.00180 at 430 nm, which agrees with the expectation. 
\begin{table}[!htp]
\caption{The refractive indices for LAB measured by V-prism refractometer, temperature is $19.2\pm1.0\,^\circ$C, humidity is 48\%, uncertainty $U=0.00005$ ($k=3$).\label{refractivity} }
\centering
\begin{tabular}{ccc}
\hline
Spectrum line & Wavelength (nm) & Refractive index \\
\hline
F line        & 486.1           & 1.49101          \\
e line        & 546.1           & 1.48559          \\
d line        & 587.6           & 1.48295          \\
D line        & 589.3           & 1.48277          \\
C line        & 656.3           & 1.47959          \\
\hline
\end{tabular}
\end{table}

Many organic liquids, such as benzene, toluene, and n-Alkanes, have large depolarization ratios\cite{Coumou,Nagai,Wahid,Kerker}. Therefore the depolarization ratio of LAB is most likely not negligible. According to Eq. (\ref{delta}) the depolarization ratio $\delta$ can be obtained based on the parallel and perpendicular scattering intensities at 90$^\circ$. Therefore, it can be derived from the recent scattering experiment\cite{Wurm}, in which these intensities have been measured at 4 angles ($\theta$ = 75$^\circ$, 90$^\circ$, 105$^\circ$ and 120$^\circ$) in the scattering plane. A Roithner Lasertechnik 430-06U LED emitting at 430 nm has been used as an unpolarized light source. The depolarization ratio of LAB at 430 nm can be derived from the published data. Based on these values, the Rayleigh scattering length can be calculated by the Einstein-Smoluchowski-Cabannes formula. 

In the scattering experiment\cite{Wurm}, the measurable quantity corrected by Monte Carlo simulations is $Q(\theta)$, which can be divided into two parts in the scattering plane.
\begin{equation}
Q(\theta)=Q_\parallel(\theta)+Q_\perp(\theta),
\end{equation}
where $Q_\parallel$ and $Q_\perp$ are the parallel and perpendicular parts of $Q(\theta)$. The angular distributions of $Q_\parallel$ and $Q_\perp$ are
\begin{eqnarray}
Q_\parallel(\theta) & = & \frac{\cos^2\theta}{2}Q_\mathrm{an}+\frac{1}{2}Q_\mathrm{is}, \nonumber \\
Q_\perp(\theta) & = & \frac{1}{2}Q_\mathrm{an}+\frac{1}{2}Q_\mathrm{is},
\label{Q}
\end{eqnarray}
where $Q_\mathrm{an}$ and $Q_\mathrm{is}$ are the fitting parameters. The analogy between Eq. (\ref{aniso}) and (\ref{Q}) indicates the anisotropy of Rayleigh scattering in LAB. It can be shown that $Q(\theta)=4\pi\beta(\theta)$. The depolarization ratio can be written as 
\begin{equation}
\delta=\frac{Q_\mathrm{is}}{Q_\mathrm{an}+Q_\mathrm{is}}.
\label{Qdelta}
\end{equation}
Moreover, the Rayleigh scattering length of LAB at 430 nm can also be obtained by
\begin{equation}
l_\mathrm{Ray}=\left(\frac{2}{3}Q_\mathrm{an}+Q_\mathrm{is}\right)^{-1}.
\label{RayQ}
\end{equation}
The factors in front of $Q_\mathrm{an}$ and $Q_\mathrm{is}$ are originating from the integral in Eq.~(\ref{Rayintegral}) over the solid angle. The Rayleigh scattering length given by Eq. (\ref{ESC}) and Eq. (\ref{RayQ}) can be cross-checked by each other. 

Based on the values and uncertainties for $Q_\mathrm{an}$ and $Q_\mathrm{is}$ given for 430 nm in the publication, the depolarization ratio and the Rayleigh scattering length of LAB can be determined. The results are listed in Table~\ref{table}. The results obtained for three different brands of LAB are in good mutual agreement within the margin of the error. The average depolarization ratio is 0.31$\pm$0.04, and the average Rayleigh scattering length is 27.0$\pm$2.3 m. 
\begin{table}[!htp]
\caption{The fitting parameters $Q_\mathrm{an}$ and $Q_\mathrm{is}$, the depolarization ratio $\delta$ and the Rayleigh scattering length $l_\mathrm{Ray}$ of LAB at 430 nm derived from the published experiment data in Ref.~\cite{Wurm}.\label{table}}
\centering
\begin{tabular}{lccccccccc}
\hline
Sample              & $Q_\mathrm{is}$   & $Q_\mathrm{an}$    & $\delta$      & $l_\mathrm{Ray}$ \\
                    &10$^{-4}$cm$^{-1}$ & 10$^{-4}$cm$^{-1}$ &               & m                \\
\hline                                                                       
LAB$_\mathrm{P500}$ & 1.33$\pm$0.09     & 3.32$\pm$0.36      & 0.29$\pm$0.03 & 28.2$\pm$2.1     \\
LAB$_\mathrm{P550}$ & 1.65$\pm$0.10     & 3.29$\pm$0.42      & 0.33$\pm$0.03 & 26.0$\pm$2.0     \\
LAB$_\mathrm{550Q}$ & 1.51$\pm$0.13     & 3.33$\pm$0.38      & 0.31$\pm$0.03 & 26.8$\pm$2.1     \\
\hline
\end{tabular}
\end{table}

Using the Einstein-Smoluchowski-Cabannes formula, the Rayleigh scattering length can be determined to 28.5$\pm$2.3 m at 23$^\circ$C, consistent with the above values. The precision of this result is limited by the uncertainty of the depolarization ratio, which could be improved much by a relative measurement\cite{Farinato}. The results are summarized in Table~\ref{compare}. 
\begin{table*}[!htp]
\caption{Summary results of LAB.\label{compare}}
\centering
\begin{tabular}{lccccccc}
\hline
Sample & $\lambda$   & T         & $n$              & $\kappa_T$                 & $\delta$(Avg.)      & $l_\mathrm{Ray}$(Theo.) & $l_\mathrm{Ray}$(Avg. of Exp.)\\
       & nm          & $^\circ$C &                  & 10$^{-10}$Pa$^{-1}$        &                     & m                       & m                     \\
\hline                                                           
LAB    & 430         & 23        & 1.498$\pm$0.004  & 7.743$\pm$0.035            & 0.31$\pm$0.04       & 28.5$\pm$2.3            & 27.0$\pm$2.3          \\
\hline
\end{tabular}
\end{table*}

\section{Discussions and conclusions\label{sec4}}
For liquids featuring a negligible depolarization ratio, as for instance water and liquid noble gases, the Rayleigh scattering lengths can be determined by the Einstein-Smoluchowski theory, without an explicit measurement of the scattered light amplitude. This approach is widely used in noble liquids detectors for solar neutrinos and dark matter searches\cite{Seidel} as well as for the water in Cherenkov detectors for atmospheric neutrinos. However, the depolarization ratio can not be neglected in the case of organic liquids like LAB. The non-zero depolarization ratio allows for both isotropic and anisotropic contributions to the Rayleigh scattering amplitude of LAB. 

In this work, the Rayleigh scattering length of LAB has been obtained not only by the use of the Einstein-Smoluchowski-Cabannes formula but also by re-evaluating the data from previous scattering experiments. Both results are in good agreement within the margin of error. The resulting Rayleigh scattering length of LAB at 430 nm is about 30 m. It is therefore substantially shorter than the value of 40 m reported earlier\cite{Wurm}. This difference is mainly caused by the fact that only part of the observed scattering amplitude following a $(1+\cos^2\theta)$ dependence was interpreted as Rayleigh scattering in the previous study, while the remaining part was ascribed to absorption/re-emission processes. However, the absorption of LAB is expected to be very weak for wavelengths longer than 340 nm\cite{Xiao}. Thus for photons with wavelength of 430 nm the absorption/re-emission process in pure LAB should be negligible compared to Rayleigh scattering process.  
 
Based on an attenuation length of 20 m in LAB at 430 nm\cite{Gao}, an absorption length of 60 m can be obtained for a Rayleigh scattering length of 30 m\cite{Wurm}. Simulations have shown that the energy resolution of the central detector of JUNO can reach $3\%/\sqrt{E_{e^+}[\mathrm{MeV}]}$ under these circumstances, thus satisfying the requirements for a measurement of the mass hierarchy in JUNO\cite{Wang}. The central detector of JUNO will be by far the largest neutrino detector using a target of organic liquid scintillator, surpassing all present-day liquid-scintillator detectors by at least a factor 20 in mass. Our study can also be useful for future large scale detectors upwards of 50 kilotons, such as the LENA experiment\cite{WurmLENA}.

%\subsection{}
%\subsubsection{}

% If in two-column mode, this environment will change to single-column format so that long equations can be displayed. 
%\begin{widetext}
%$$\mbox{put long equation here}$$
%\end{widetext}

% Figures should be put into the text as floats. 
% Use the graphics or graphicx packages (distributed with LaTeX2e). EPSFig is no longer fully supported.
% Put the label that you will use with \ref{} command in the braces of the \label{} command.
%
% \begin{figure}
% \includegraphics{}% % Important NOTE: Please make certain your figures do not include local directory paths. ex. "c:\file\sub\fig1.eps"
% \end{figure}

% \begin{table}
% \end{tabular}
% \end{table}

\begin{acknowledgments}
% Put your acknowledgments here.
This work has been supported by the Major Program of the National Natural Science Foundation of China (Grant No. 11390381), the Strategic Priority Research Program of the Chinese Academy of Sciences (Grant No. XDA10010500), the 985 project of Wuhan University (Grant No. 202273344).
\end{acknowledgments}

%\bibliography{your-bib-file}
% Run this once to generate your BBL file. Then copy the contents of your BBL file into your main latex file, commenting out "\bibliography"

\end{document}